\documentclass{webofc}
\usepackage[varg]{txfonts}   
\begin{document}
\title{Heavy quark flow as better probes of QGP properties}
%
%

\author{\firstname{Zi-Wei} \lastname{Lin}\inst{1,2}\fnsep\thanks{\email{linz@ecu.edu}}
\and
        \firstname{Hanlin}
        \lastname{Li}\inst{3,4}\fnsep\thanks{\email{lihl@wust.edu.cn}}
\and
        \firstname{Fuqiang}
        \lastname{Wang}\inst{4,5}\fnsep\thanks{\email{fqwang@purdue.edu}}
}
\institute{
Key Laboratory of Quarks and Lepton Physics (MOE) and Institute of Particle Physics,
Central China Normal University, Wuhan 430079, China
\and
Department of Physics, East Carolina University, Greenville, North
Carolina 27858, USA
\and
College of Science, Wuhan University of Science and Technology, Wuhan,
Hubei 430065, China
\and
Department of Physics and Astronomy, Purdue University, West
Lafayette, IN 47907, USA
\and
School of Science, Huzhou University, Huzhou, Zhejiang 313000, China
}
\abstract{
In earlier studies we have proposed that most parton $v_2$ comes from the
anisotropic escape of partons, not from the hydrodynamic flow, even
for semi-central Au+Au collisions at $\sqrt {s_{NN}}=200$ GeV. 
Here we study the flavor dependence of this escape mechanism with a
multi-phase transport model. 
In contrast to naive expectations, we find that the charm $v_2$ is much
more sensitive to the hydrodynamic flow than the lighter quark $v_2$,
and the fraction of $v_2$ from the escape mechanism
decreases strongly with the quark mass for large collision
systems. We also find that the light quark collective flow is
essential for the charm quark $v_2$. Our finding thus suggests that
heavy quark flows are better probes of the quark-gluon-plasma
properties than light quark flows. 
}
\maketitle
\section{Introduction}
Azimuthal anisotropies in heavy ion collisions, such as the elliptic
flow $v_2$, are important tools for the study of the properties of the
quark-gluon plasma (QGP). Recent studies with parton transport models suggest 
\cite{He:2015hfa,Lin:2015ucn} that most parton $v_2$ comes from the
anisotropic escape of partons, not from the hydrodynamic flow, even
for semi-central Au+Au collisions at $\sqrt {s_{NN}}=200$ GeV. 
This escape mechanism converts the spatial anisotropy in the
overlap volume very efficiently into azimuthal anisotropies in final state
particles, even though the parton cross section and
average number of collisions of each parton are small
\cite{He:2015hfa,Lin:2015ucn,Adare:2015ctn,Koop:2015trj}. 
Thus it naturally explains the similar azimuthal
anisotropies observed in small and large collision systems. 
However, it poses a challenge to the current perfect-fluid paradigm
for heavy ion collisions, at least when the collision system and/or
energy are not very large.

Our earlier studies \cite{He:2015hfa,Lin:2015ucn} 
looked at all quarks (regardless of their flavors) and 
only investigated d+Au and Au+Au systems at $200A$ GeV.
Here we study the flavor dependence of the parton escape mechanism,
especially the charm quarks \cite{Li:2016hbf,work}. 
As in the earlier studies, we use the string melting version of a
multi-phase transport (AMPT) model \cite{Lin:2004en} with the same
parameters , which can reasonably describe the experimental data for
the bulk matter for high energy heavy ion collisions \cite{Lin:2014tya}. 
We follow the entire evolution history of quarks of different flavors in
AMPT and then analyze the developments of light (u and d quarks),
strange, and charm $v_2$ in three systems: p+Pb collisions at $5A$ TeV
and impact parameter $b\!=\!0$ fm, Au+Au collisions at $200A$ GeV and $b\!\in\!
(6.6,8.1)$ fm, and Pb+Pb collisions at $2.76A$ TeV and $b\!=\!8$ fm. 
Note that only the results for the Pb+Pb collisions are shown in the
figures here, and the results of quarks and antiquarks of the same 
flavor have been combined. 

\section{Results}

To follow the parton collision history in the AMPT model, we define
$N_{coll}$ as the number of collisions suffered by a parton.
At any given $N_{coll}$ value, we study three groups of quarks of a
given flavor: freezeout partons (partons that freeze out 
after exactly $N_{coll}$ collisions), 
non-freezeout partons (partons with more than $N_{coll}$ collisions),
and all (active) partons (sum of the previous two groups). 

\begin{figure}[h]
\centering
\includegraphics[height=4.2cm]{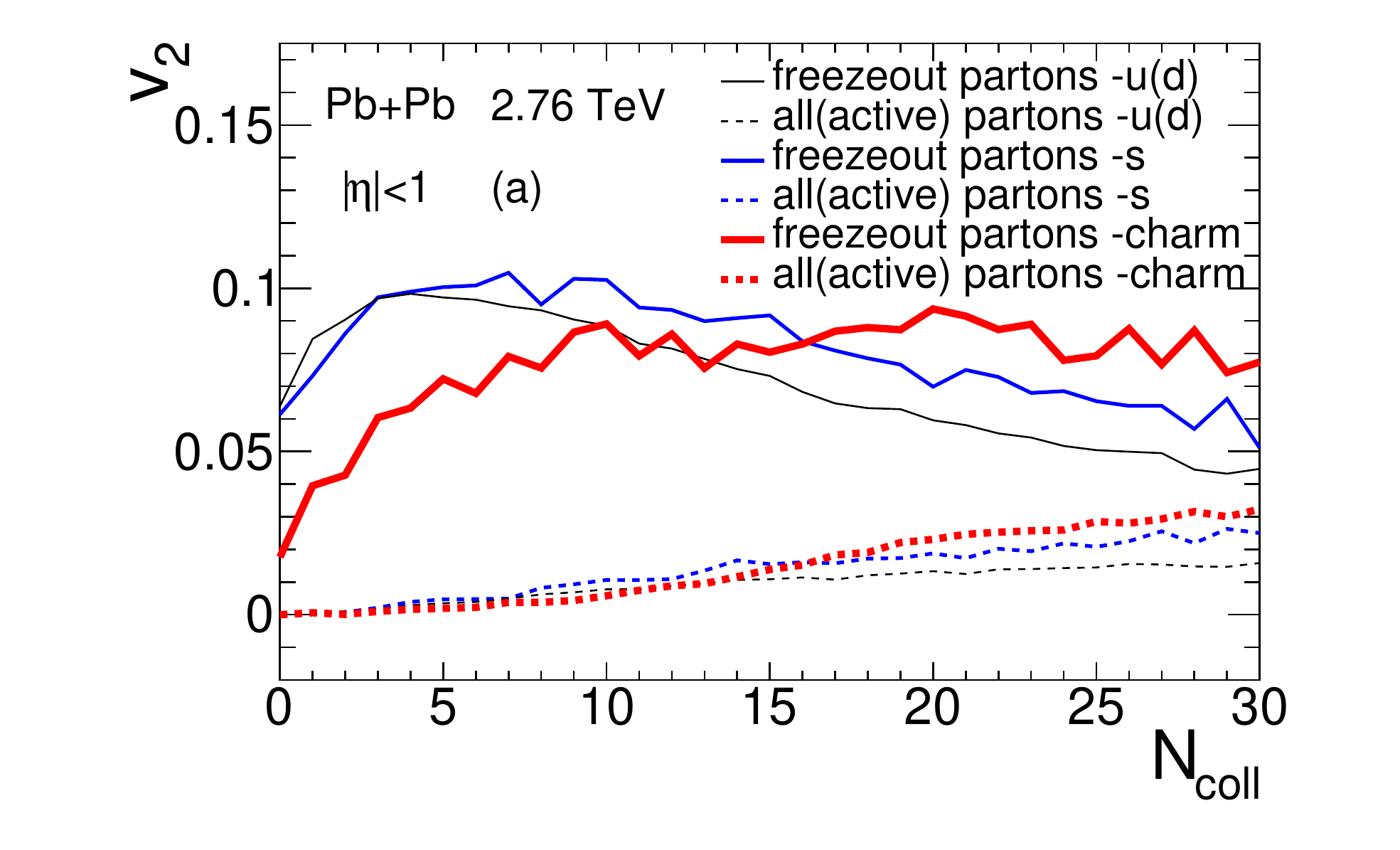}
\includegraphics[height=4.2cm]{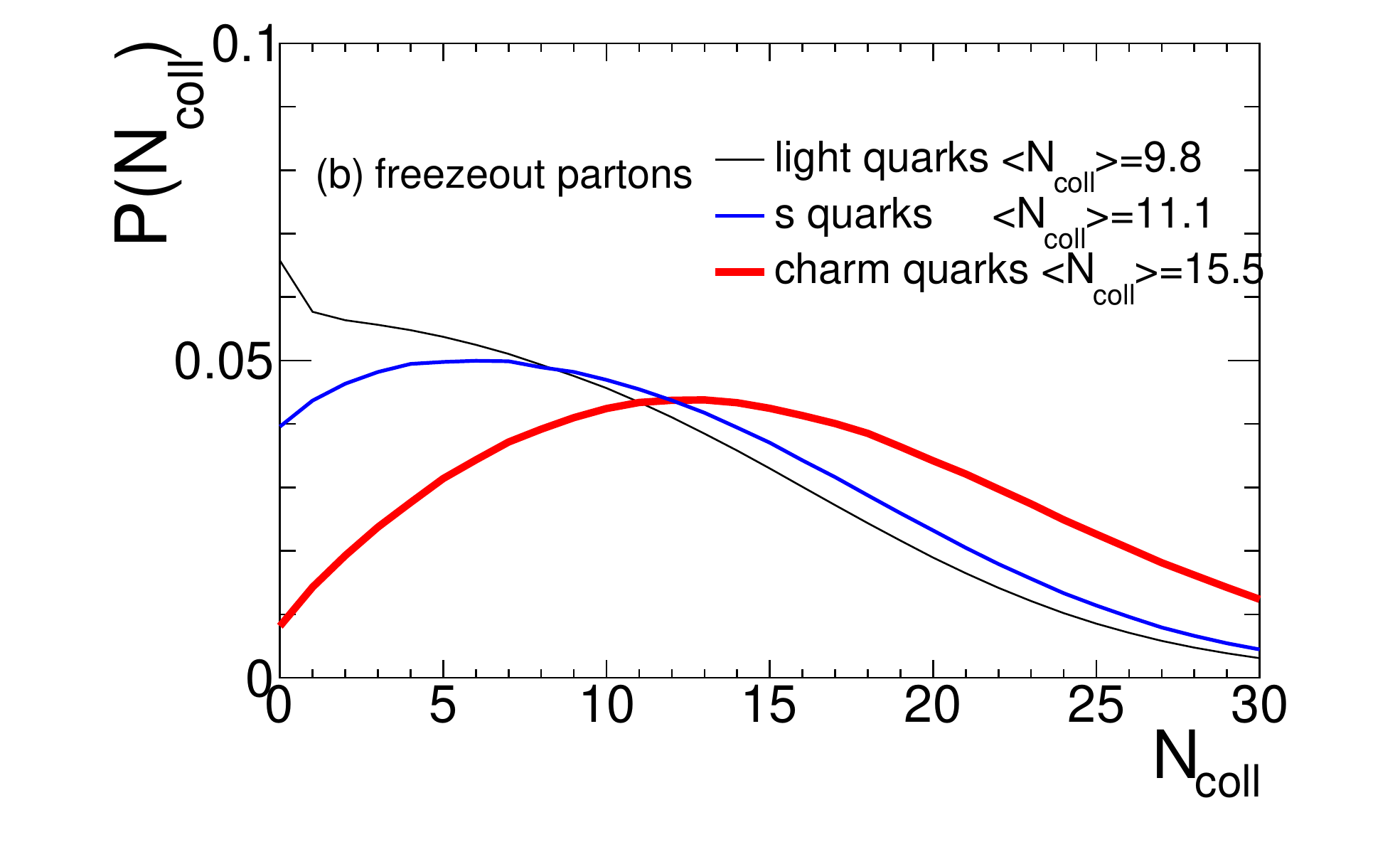}
\vspace{-0.6cm}
\caption{
AMPT simulations of Pb+Pb collisions at 2.76A TeV and $b\!=\!8$ fm:
(a) $v_2$ of light, strange, and charm quarks within $|\eta|<1$  as
functions of $N_{coll}$, where solid curves represent freezeout parton 
and dashed curves represent all (active) partons;
(b) normalized $N_{coll}$-distributions of different quark flavors.
}
\label{fig1}       
\vspace{-0.3cm}
\end{figure}

Figure~\ref{fig1}a shows the $v_2$ of light (black), strange (blue),
and charm (red) quarks within $|\eta|<1$  as functions of the number
of collisions $N_{coll}$ for $2.76A$ TeV Pb+Pb collisions at $b\!=\!8$
fm from AMPT simulations. We see a clear mass ordering in that 
$v_{2,u/d} > v_{2,s} > v_{2,c}$ at low $N_{coll}$ but the opposite at
high $N_{coll}$, indicating that charm quarks need more scatterings to 
generate their $v_2$.
The normalized $N_{coll}$ distribution and the average number of
collisions of each quark flavor (of all pseudorapidities) are shown in
Fig.~\ref{fig1}b, where we see that charm quarks typically have more
collisions than lighter quarks. 
This should be related to the initial spatial and momentum
distributions, which are different for each quark flavor. For example,
we find that, in comparison with light quarks, a bigger fraction of
charm quarks is produced in the inner region of the overlap volume, 
consistent with the hard production nature of charm quarks and their
scaling with binary collisions. 
In addition, we find that the above features in Fig.1 are true
for all three collision systems in our study. 

For partons that freeze out (i.e. hadronize) at $N_{coll}=0$, 
their finite positive $v_2$, as shown in Fig.~\ref{fig1}a, 
is due to the fact that it is easier to avoid collisions along the
impact parameter direction in the transverse plane. 
As these partons have not been affected by the collective flow, 
their $v_2$ comes purely from the anisotropic escape probability, 
an interaction-induced response to the anisotropic geometrical shape
that we named the escape mechanism. 
For partons that freeze out at $N_{coll} \neq 0$, however, 
their $v_2$ comes partly from the interaction-induced response to
geometry and partly from the collective flow that is also anisotropic.
In order to identify the contribution from the escape mechanism, 
we have designed the azimuth-randomized test to remove the
collective flow \cite{He:2015hfa,Lin:2015ucn}, where we randomize the azimuth angle of
each of the two final state partons after every parton scattering. 
As a result, anisotropic flows in the  azimuth-randomized simulations 
are generated only by the escape mechanism. 

\begin{figure}[h]
\centering
\begin{minipage}{4.6cm}
\centering
 \includegraphics[height=3.cm]{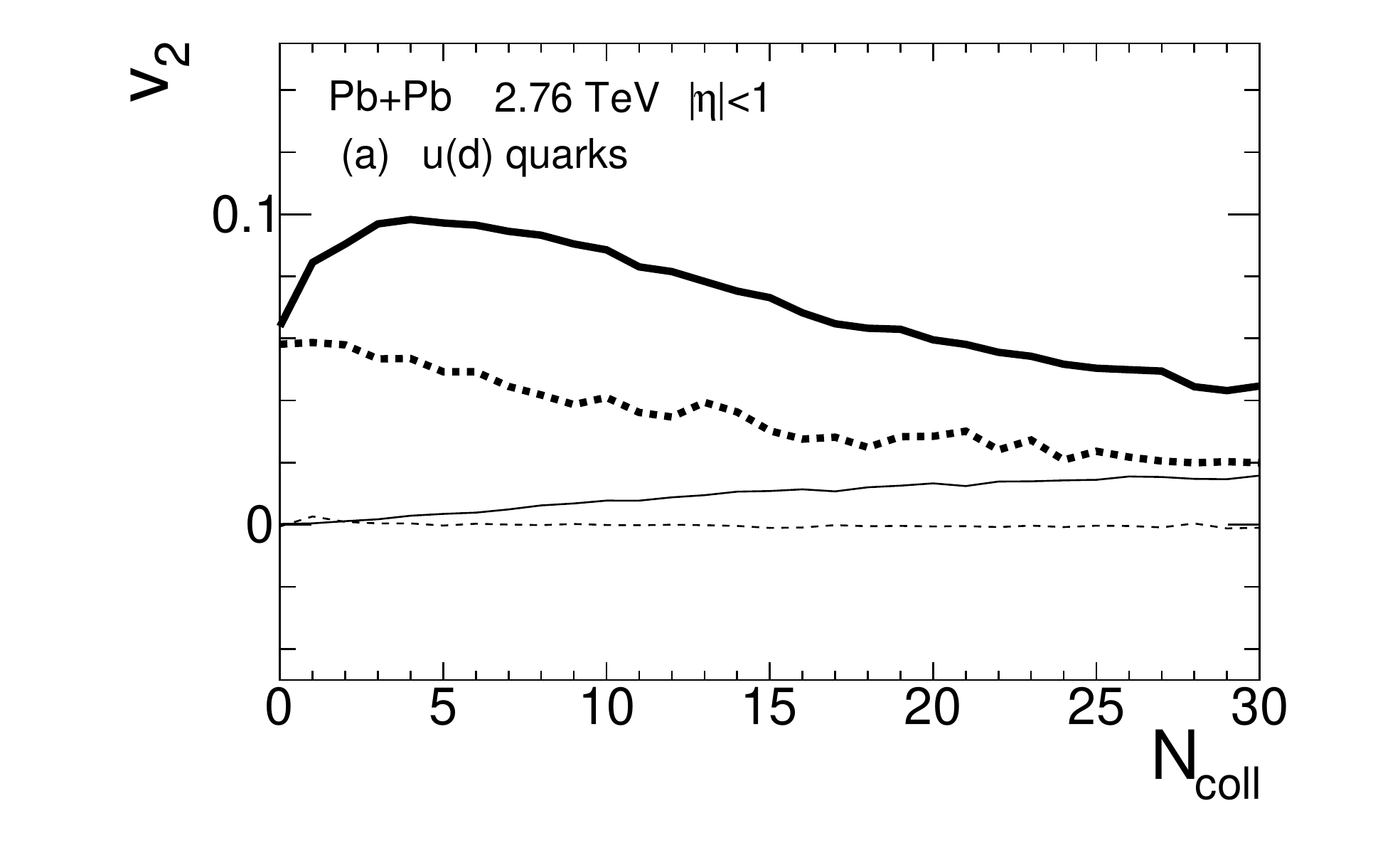}
 \end{minipage} 
\begin{minipage}{4.6cm}
 \includegraphics[height=3.cm]{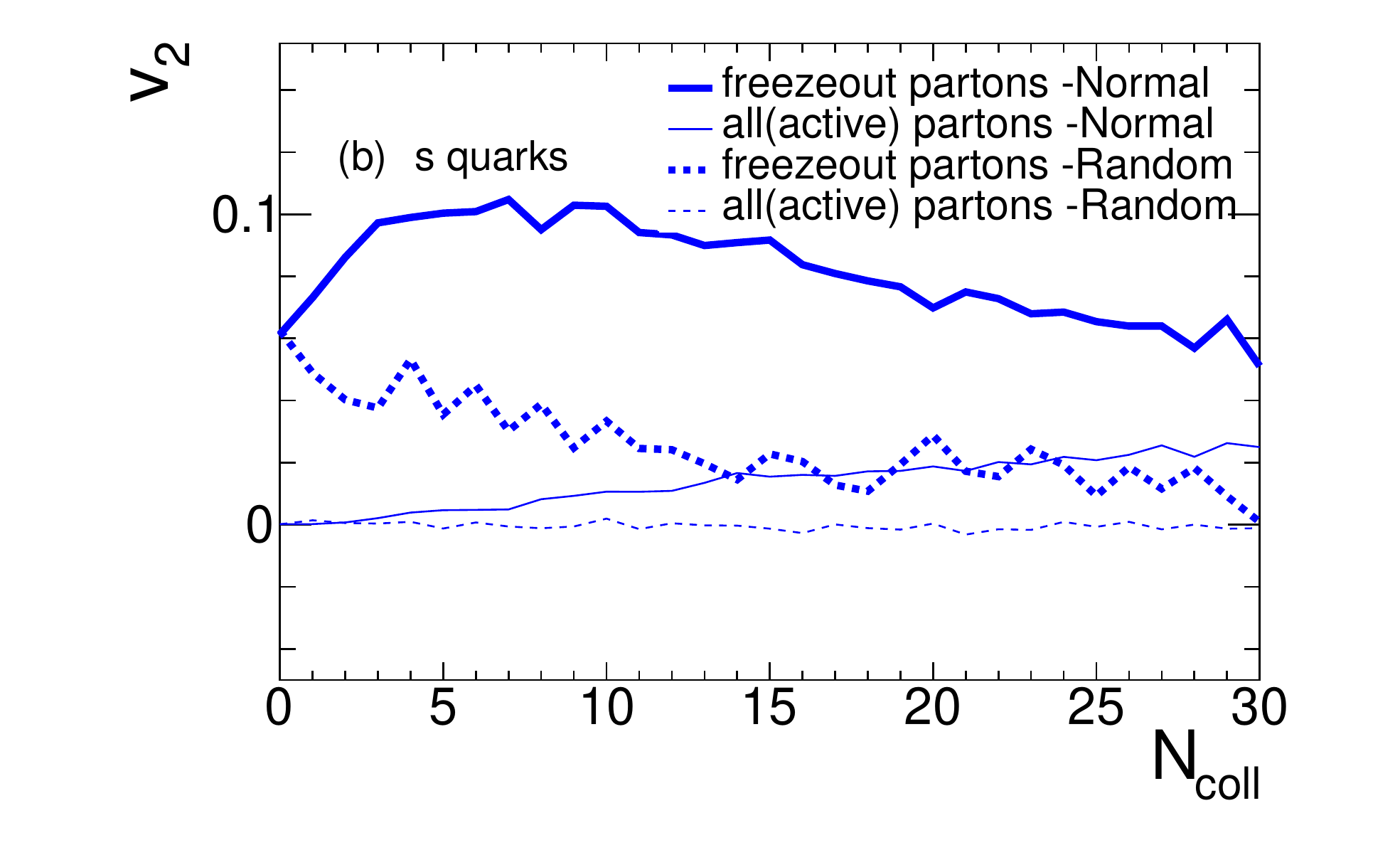}
 \end{minipage}
\begin{minipage}{4.5cm}
 \includegraphics[height=3.cm]{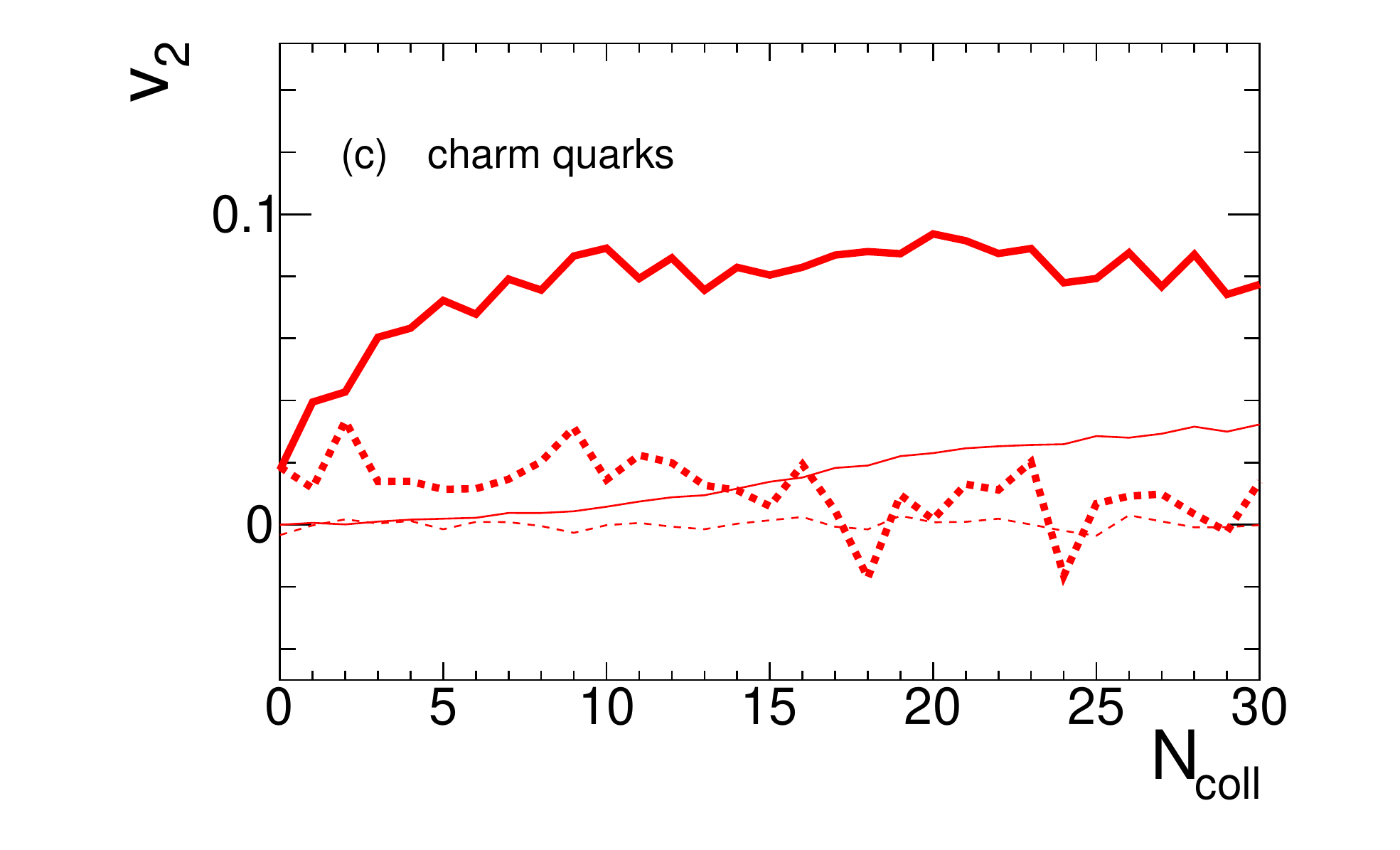}
 \end{minipage}
\vspace{-0.3cm}
\caption{
The $v_2$ of (a) light, (b) strange, and (c) charm quarks within
$|\eta|<1$ as functions of
$N_{coll}$ for Pb+Pb collisions at 2.76A TeV and $b\!=\!8$ fm.}
\label{fig2}       
\vspace{-0.3cm}
\end{figure}

We show in Fig.~\ref{fig2} the $v_2$ as functions of
$N_{coll}$ for light, strange, and charm quarks within $|\eta|<1$ 
from both normal AMPT simulations (solid curves) 
and azimuth-randomized AMPT simulations (dashed curves) 
of Pb+Pb collisions. 
Note that the average parton $v_2$ is the freezeout
$v_2$ shown here summed with the weight given by the
$N_{coll}$-distribution shown in Fig.~\ref{fig1}b. 
First we see that the $v_2$ results from azimuth-randomized calculations 
are finite, although they are mostly lower than that from normal
calculations due to the lack of the anisotropic collective flow. 
We also see that the reduction of $v_2$ going from normal to
azimuth-randomized calculations is more obvious for heavier quarks, 
indicating that a smaller fraction of $v_2$ comes from the escape
mechanism for heavier quarks. 

\begin{figure}[h]
\centering
\includegraphics[height=4.1cm]{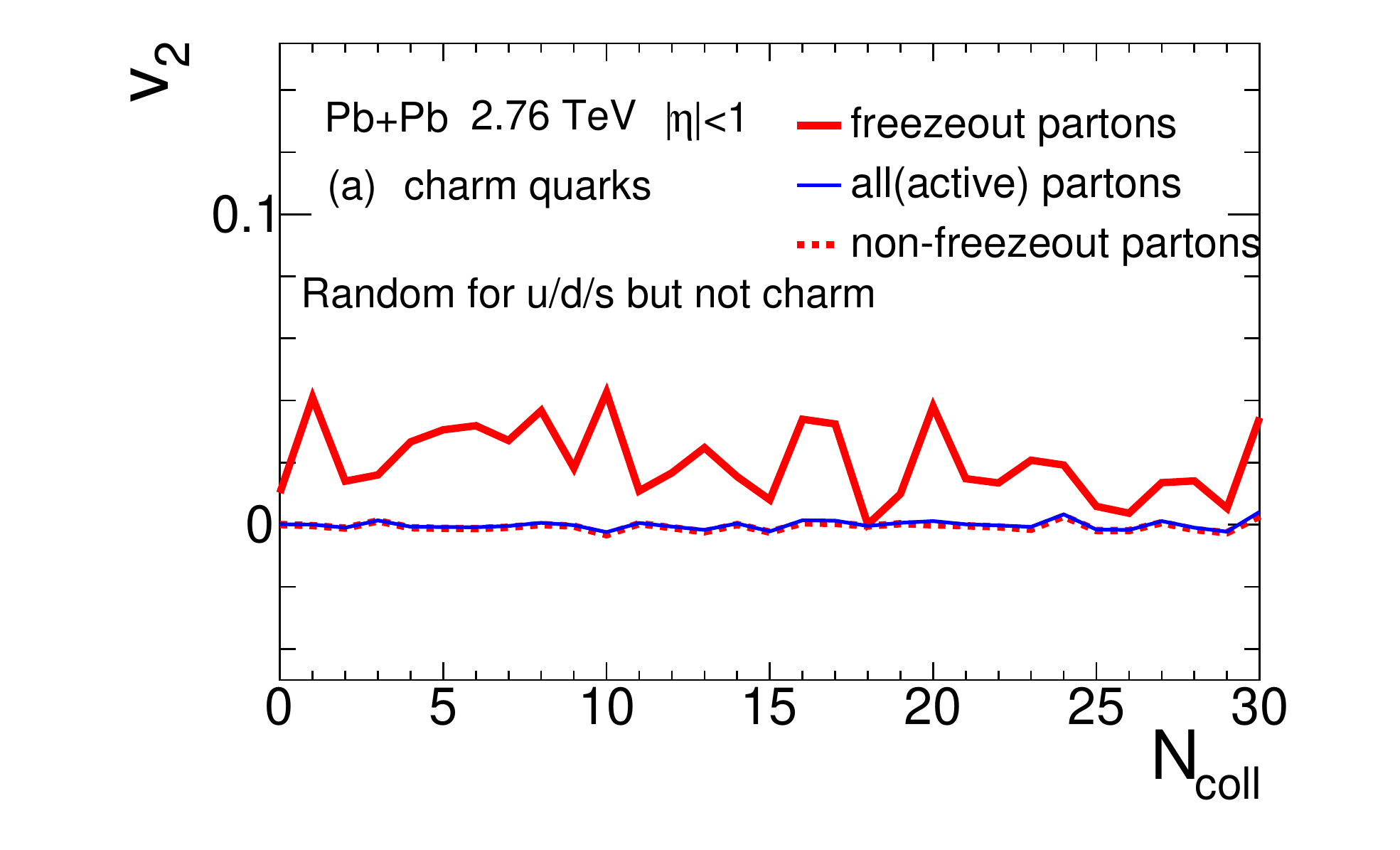}
\includegraphics[height=4.2cm]{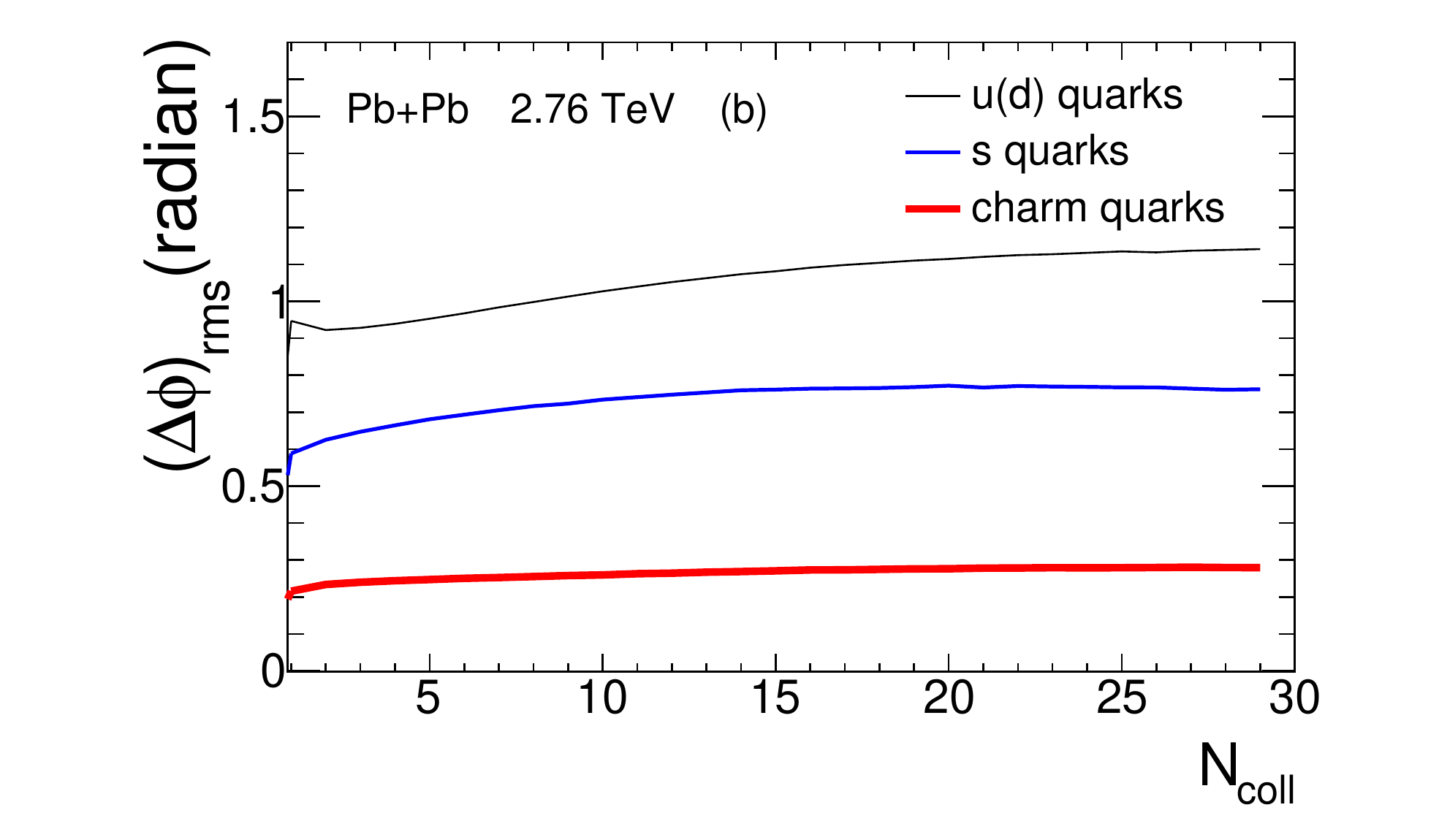}
\vspace{-0.5cm}
\caption{
AMPT results of 2.76A TeV Pb+Pb collisions at $b\!=\!8$ fm:  
(a) $v_2$ of three charm quark groups within $|\eta|<1$
as functions of $N_{coll}$ in simulations where u/d/s (but not charm) 
(anti)quarks are azimuth-randomized; 
(b) the rms change of azimuth due to the $N_{coll}\!-\!th$
collision for different quark flavors in normal simulations.
}
\label{fig3}       
\vspace{-0.3cm}
\end{figure}

Another interesting test is to do azimuth-randomized calculations 
only for u/d/s quarks and antiquarks but not on charm, i.e., 
only charm quarks and antiquarks are allowed to keep their collective
flow developed in the parton cascade. 
Results from this test are shown in Fig.~\ref{fig3}a for charm quarks
within $|\eta|<1$ in Pb+Pb collisions, where we see that the freezeout 
charm $v_2$ is much reduced when compared to the normal charm
$v_2$ (thick solid curve in Fig~\ref{fig2}c) and is similar to the
freezeout charm $v_2$ when all quark flavors are azimuth-randomized (thick
dashed curve in Fig~\ref{fig2}c). This means that charm quarks cannot
develop a significant $v_2$ without the light quark collective flow.

To further understand the flavor dependence, 
we have also looked at the effect of each parton collision on the
azimuth of different flavors. 
Figure~\ref{fig3}b shows the root-mean-square (rms) change of
the azimuth angle, in radian, as a function of $N_{coll}$ for
different quark flavors (of all pseudorapidities) in Pb+Pb
collisions. There is a clear mass ordering in that  
the average azimuth change is much smaller for heavier quarks, 
consistent with the expectation that it is more difficult to deflect a
heavier quark in the parton cascade. Note that AMPT uses 
the same light quark mass as PYTHIA \cite{Sjostrand:1993yb}: 
$m_u=5.6, m_d=9.9$, $m_s = 199$ MeV/$c^2$, and we use $m_c=1.2$
GeV/$c^2$ for charm in this study. 

Table~\ref{tab} lists the ratios of the $\langle v_2 \rangle$ 
from azimuth-randomized AMPT over that from normal AMPT, 
where $\langle v_2 \rangle$ represents the $v_2$ averaged over all
partons within $|\eta|<1$, in four different collision systems at RHIC
and LHC energies. This ratio represents the fraction of $v_2$ that
comes from the escape mechanism, i.e., from the 
interaction-induced response to the anisotropic geometry. 
Note that the d+Au result comes from our earlier study 
on quarks of all flavors \cite{He:2015hfa}. 
We can see that the escape contribution decreases with the quark mass,
more strongly for larger systems. It also decreases with the collision
energy and/or system size, which is expected. 
Therefore these results show that the hydrodynamical collective flow
contributes more to the $v_2$ of heavier quarks, especially in large
systems at high energies, suggesting that heavy quark flows can better
reflect the properties of the quark-gluon plasma. We note that similar
claims have been made in studies from other points of view 
\cite{Esha:2016svw,Greco:2017rro}.

\begin{table}
\centering
\caption{Ratio of the averaged $v_2$ from azimuth-randomized simulations
  over that from normal simulations, which represents the fraction
  from the escape mechanism.}
\label{tab}       
\begin{tabular}{cccccc}
\hline
 & d+Au at 200 GeV & p+Pb at 5 TeV & Au+Au at 200 GeV & Pb+Pb at 2.76 TeV \\
 & ($b\!=\!0$ fm) & ($b\!=\!0$ fm) & ($b\!=\!6.6-8.1$ fm) 
& ($b\!=\!8$ fm)  \\\hline 
u/d & 93\% (all flavors) & 72.9\% & 65.6\% & 42.5\% \\
s &  & 59.1\% & 47.4\% & 26.5\% \\
c &  & 56.8\% & 21.8\% & 8.5\% \\ \hline
\end{tabular}
\vspace{-0.3cm}
\end{table}

\section{Summary}

We have followed the complete parton collision history to 
study the $v_2$ of light, strange and charm quarks 
in small and large collision systems at both RHIC and LHC energies 
using the string melting version of the AMPT model. 
We find that the fraction from the interaction-induced response to the 
anisotropic spatial geometry (the escape mechanism) decreases not only 
with the system size and collision energy but also with the quark
mass (especially for large systems at high energies).
The escape mechanism is no longer dominant for the light quark average
$v_2$ in semi-central Pb+Pb collisions at 2.76 TeV but its
contribution is stil significant.  
On the other hand, most of the charm quark $v_2$ comes from the 
hydrodynamical collective flow for the large systems covered in this
study. 
We also find that the collective flow of light quarks
is essential for the generation of charm quark $v_2$. 
These results indicate that heavy quark flows are better probes of QGP 
properties than light quark flows. 

This work is supported in part by the NSFC of China under Grants
Nos. 11628508, 11647306, and US Department of Energy
Grant No. DE-SC0012910. HL acknowledges the financial support from the
China Scholarship Council. 
%

\begin{thebibliography}{}
\bibitem{He:2015hfa} 
  L.~He, T.~Edmonds, Z.~W.~Lin, F.~Liu, D.~Molnar and F.~Wang,
  Phys.\ Lett.\ B {\bf 753}, 506 (2016).
\bibitem{Lin:2015ucn} 
  Z.~W.~Lin, L.~He, T.~Edmonds, F.~Liu, D.~Molnar and F.~Wang,
  Nucl.\ Phys.\ A {\bf 956}, 316 (2016).
\bibitem{Adare:2015ctn} 
  A.~Adare {\it et al.} [PHENIX Collaboration],
  Phys.\ Rev.\ Lett.\  {\bf 115}, 142301 (2015).
\bibitem{Koop:2015trj} 
  J.~D.~Orjuela Koop, R.~Belmont, P.~Yin and J.~L.~Nagle,
  Phys.\ Rev.\ C {\bf 93}, 044910 (2016).
\bibitem{Li:2016hbf} 
  H.~Li, Z.~W.~Lin and F.~Wang,
  J.\ Phys.\ Conf.\ Ser.\  {\bf 779}, 012063 (2017).
\bibitem{work} 
H.~L. Li, Z.~W. Lin, and F. Wang, in preparation.
\bibitem{Lin:2004en} 
  Z.~W.~Lin, C.~M.~Ko, B.~A.~Li, B.~Zhang and S.~Pal,
  Phys.\ Rev.\ C {\bf 72}, 064901 (2005).
\bibitem{Lin:2014tya} 
  Z.~W.~Lin,
  Phys.\ Rev.\ C {\bf 90}, 014904 (2014).
\bibitem{Sjostrand:1993yb} 
  T.~Sjostrand,
  Comput.\ Phys.\ Commun.\  {\bf 82}, 74 (1994).
\bibitem{Esha:2016svw} 
  R.~Esha, M.~Nasim and H.~Z.~Huang,
  J.\ Phys.\ G {\bf 44}, 045109 (2017).
\bibitem{Greco:2017rro} 
  V.~Greco,
  Nucl.\ Phys.\ A {\bf 967}, 200 (2017).
\end{thebibliography}
%
%

\end{document}